\documentclass[10pt,twocolumn,letterpaper]{article}

\usepackage{iccv}
\usepackage{times}
\usepackage{graphicx}
\usepackage{amsmath}
\usepackage{amssymb}

\usepackage[pagebackref=true,breaklinks=true,letterpaper=true,colorlinks,bookmarks=false]{hyperref}

\iccvfinalcopy 

\ificcvfinal\fi

\begin{document}

\title{TeliNet: Classifying CT scan images for COVID-19 diagnosis}

\author{Mohammad Nayeem Teli\\
Department of Computer Science\\
University of Maryland\\
College Park, MD 20742\\
{\tt\small nayeem@umd.edu}
}

\maketitle
\ificcvfinal\thispagestyle{empty}\fi

\begin{abstract}
   COVID-19 has led to hundreds of millions of cases and millions of deaths worldwide since its onset. The fight against this pandemic is on-going on multiple fronts. While vaccinations are picking up speed, there are still billions of unvaccinated people. In this fight against the virus, diagnosis of the disease and isolation of the patients to prevent any spread play a huge role. Machine Learning approaches have assisted in the diagnosis of COVID-19 cases by analyzing chest X-rays and CT-scan images of patients. To push algorithm development and research in this direction of radiological diagnosis, a challenge to classify CT-scan series was organized in conjunction with ICCV, 2021. In this research we present a simple and shallow Convolutional Neural Network based approach, TeliNet, to classify these CT-scan images of COVID-19 patients presented as part of this competition. Our results outperform the F1 `macro' score of the competition benchmark and VGGNet approaches. Our proposed solution is also more lightweight in comparison to the other methods.
\end{abstract}

\section{Introduction}

On March 11, 2020, the World Health Organization(WHO) declared COVID-19, caused by a novel coronavirus, a global pandemic~\cite{who-covid19}. Soon after that the global lock downs and restrictions on travel brought life to a stand still as we know it. Since then there have been over 208 Million cases with over 4.3 Million deaths caused by this virus worldwide, as of the middle of August, 2021. In an effort to combat this virus our main defenses were isolation, quarantining, face masks and frequent washing of hands. Some of these approaches continue till date as we transition into in-person work and other activities.

In addition to these simple measures, frequent testing and tracing back helps identify and isolate spreaders and super spreaders of the virus. The real game changer in preventing hospitalizations and deaths in this yet to be over fight against this virus have been the vaccines. However, with so many cases, and slower rate of vaccinations the researchers have relied on many types of diagnostic techniques and tests to identify a COVID-19 positive patient.

There are three common types of tests that are carried out to identify a COVID positive patient. First type of testing,  Reverse-Transcriptase Polymerase Chain Reaction (RT-PCR) testing, identifies the presence of actual viral genetic material by taking a nasal pharyngeal swab. The second type of testing is antigen testing. It detects the presence of one of the outer proteins of the viral shell or envelope. The third type of testing is to detect the presence of antibodies inside a person that identify the presence of the virus. The RT-PCR testing is the most popular and more accurate than antigen testing. However, it is more sensitive than antigen testing and needs to be done in a laboratory setting. As a result RT-PCR testing has a turnaround time of several days. 

In addition to these tests it is a common practice to supplement diagnostic tests for detection and accurate prognosis of different diseases by applying machine learning approaches ~\cite{liuScience21,LUNDERVOLD2019102,marleen16,erickson17,kollias2018deep,lawrencejama2018,kollias2020deep,kollias2020transparent}. The most commonly used imaging diagnostic tools include, X-ray imaging, Computed Tomography (CT), ultrasound imaging and Magnetic Resonance Imaging (MRI).

Since the beginning of the Coronavirus pandemic, many machine learning approaches have been applied to help clinicians detect and diagnose COVID-19. Many of these approaches specifically use images obtained from Chest Radiographs (CXR) and chest Computed Tomography(CT)~\cite{roberts21}. In order to further push Machine Learning research for better analysis and classification of COVID-19 
diagnostic image data, a workshop is organized in conjunction with International Conference on Computer Vision (ICCV) '21 conference. In this workshop a competition is organized to classify CT-scan image series into covid and non-covid classes~\cite{kollias2021mia}. This paper presents our work on this data set in response to this COVID-19 CT scan series classification challenge.

We introduce TeliNet, a very simple and lightweight but effective Convolutional Neural Network (CNN) based approach to classify chest CT-scan series images of patients as COVID-19 positive or not. We compare our results with VGGNet-16~\cite{Simonyan15} and the benchmark classification approach presented by Dimitrios et al.~\cite{kollias2021mia} on the same data set. The results indicate that our TeliNet method performs significantly better than both these approaches. Our F1 `macro' score on the validation data set is 0.81 in comparison to the benchmark score of 0.70 and a VGGNet-16 score of 0.72. Our method is also significantly faster than VGGNet-16 when run on the same hardware. The main contributions of this paper include, 1) TeliNet, a lightweight and simple but very effective CNN based approach to classify CT-scan series, and 2) low computational requirements. Our approach was run on a MacBook Pro with 2.3 Ghz dual core i5 processor with 8 GB of RAM. This makes it easily accessible to users who may not have access to high-end computational resources.

In the next few sections we present the related work, followed by the data set used in this research. Next, we discuss the methodology, including the architectural details of TeliNet and VGGNet. Finally, we present results followed by the conclusions.

\section{Related work}
Computer vision has been a very useful technique in assisting clinicians with diagnosis. Due to the advancement of pattern recognition and classification algorithms, computer aided diagnostic tools have greatly supplemented the ability to detect different diseases~\cite{liuScience21}.
In order to classify the images obtained using X-rays, MRI and Computer Tomography, Convolution Neural Networks (CNN) have been the most popular tool~\cite{yadav2019}. Different types of CNN architectures have been used for image diagnosis since 2012, such as, AlexNet~\cite{alexnet}~\cite{Khan2017}, VGGNet~\cite{Simonyan15}, ResNet~\cite{He_2016_CVPR}, InceptionNet~\cite{inceptionnet}, etc.

Bressem et al.~\cite{bressem} studied and compared 16 different CNN architectures on two datasets,  CheXpert and COVID-19 image
data collection. Bressem et al. concluded that it is not necessary to use complex architectures and that simpler CNN's perform great in classification on both data sets. Sahlol~\cite{sahlol} combined Inception CNN as a feature extractor with Fractional-order Calculus - Marine Predators Algorithm (FO-MPA) to detect COVID-19  in images using the COVID-19 dataset available on Kaggle. Jia et al.~\cite{Jia21} used a modified MobileNet and modified ResNet architectures to classify COVID-19 and non-COVID-19 CT scan images data set~\cite{wang20}. The results indicate almost perfect accuracy on that data set. Ibrahim et al.~\cite{IBRAHIM2021104348} compared four different architectures to classify chest X-ray and CT images to detect COVID-19, pneumonia and lung cancer using a single model. Many other Convolutional Neural Network based architectures ~\cite{caio21,alam,ghaderzadeh,jameskusrini,liang,gilanie} have been similarly proposed since the onset of COVID-19 to help diagnose this disease and assist clinicians.
\section{Data set}
\begin{figure*}
\begin{center}
\includegraphics[scale=0.5]{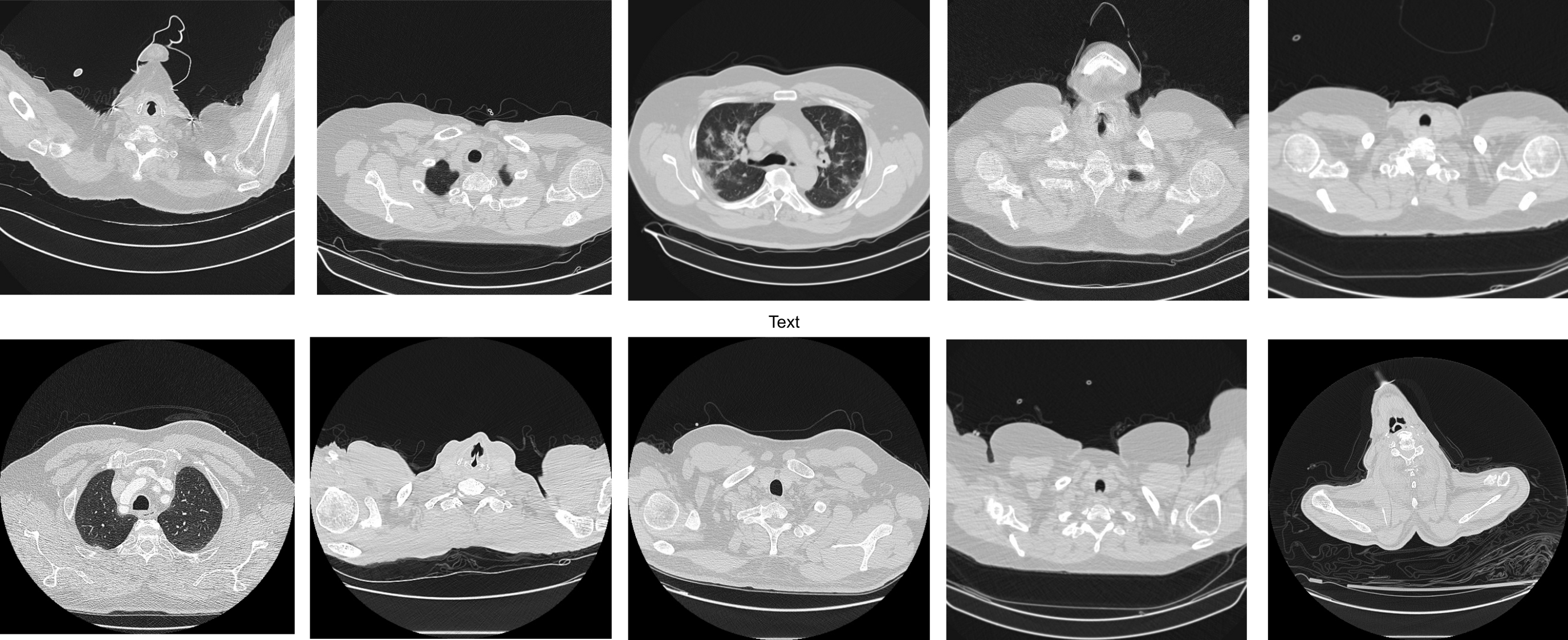}
\end{center}
   \caption{Samples of 2D CT scan slices. The top row represents covid slices and the bottom row, non-covid slices.}
\label{ctslices}
\end{figure*}

The data set used for the this research is part of ``AI-enabled Medical Images Workshop and COVID-19 Diagnosis Competition (MIA-COV19D)", organized in conjunction with the International Conference on Computer Vision (ICCV) 2021 in Montreal, Canada, October 11- 17, 2021. The data is a part of a chest CT scan database manually annotated for COVID-19 / nonCOVID-19 diagnosis. The data set consists of 5000 CT scan series' with each series consisting of 50-700 2D-CT slices. Some of the representative slices of covid and non-covid CT scan slices are presented in  Figure~\ref{ctslices}.
Each of these slices is a $512 \times 512$ pixels 2D image. There are three different subsets that are provided, training, validation and test. The training set has 335672 images belonging to 2 classes and the validation data set has 75534 images belonging to 2 classes. The test set consists of 450 folders each representing a series containing a total of 609057 2D slices. For a detailed description of this data a curious reader is encouraged to read the benchmark paper by Dimitrios et al.~\cite{kollias2021mia}. This paper describes the competition data set and the benchmark approach applied on it in great detail.

\section{Methods}
In this section we present our CNN architecture, TeliNet, and compare its results with VGGNet-16 and the benchmark results. Both, TeliNet and VGGNet-16 approaches were run on the same hardware. The following sections describe our experiments using the two architectures and their details. However, we begin with data pre-processing, followed by TeliNet and VGGNet-16 architectures' description.

\subsection{Preprocessing}
In order to better organize the data for our experiments we use data generators. These data generator inputs are applied to both TeliNet as well as VGGNet architectures. For the benchmark approach results we rely on the experiments and the results provided in the benchmark paper by Dimitrios et al.

Data generators require that the data is organized in folders representing each class. Since we have two main classes, COVID-19 and non-COVID-19 the data is stored in two folders named as such. At the top level we have two main folders train and validation. Each of these two folders contain covid and non-covid folders representing the two classes for binary classification. All images belonging to their specific class were stored in their respective folders. A separate test folder contains eight different subsets. Each of those subsets contains images for predictions with 450 folders, each representing a CT-scan series of test slices for which labels were not provided.
\subsection{TeliNet}
\begin{figure*}
\begin{center}
\includegraphics[scale=0.65]{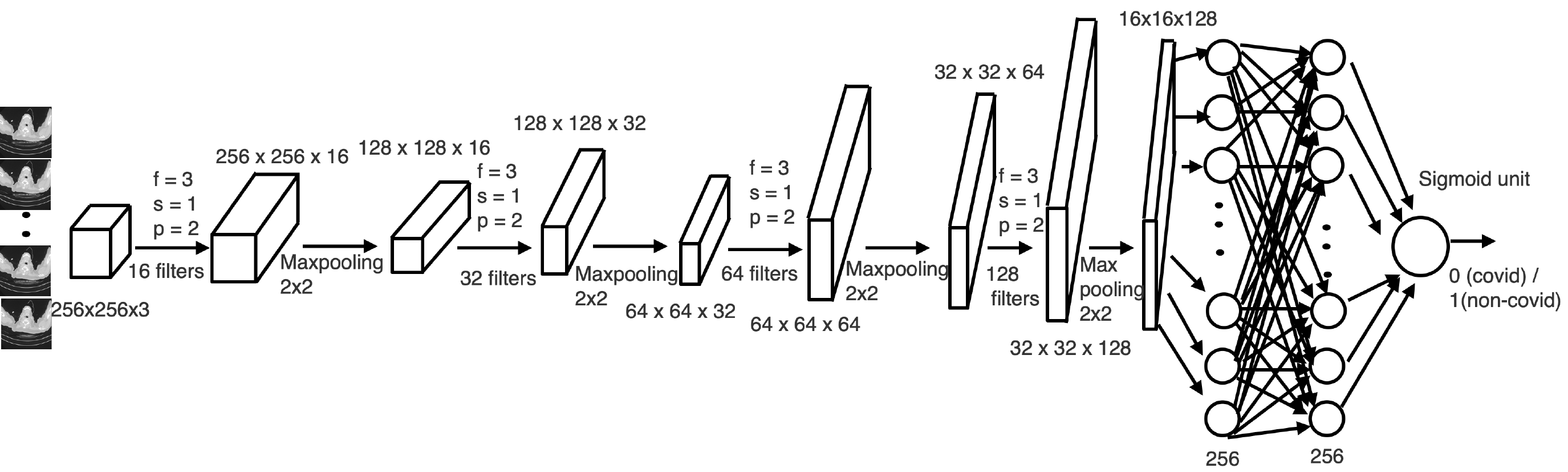}
\end{center}
   \caption{TeliNet architecture.}
\label{telinet}
\end{figure*}

\begin{figure}[t!]
\begin{center}
   \includegraphics[height=6in,width=1.2\linewidth]{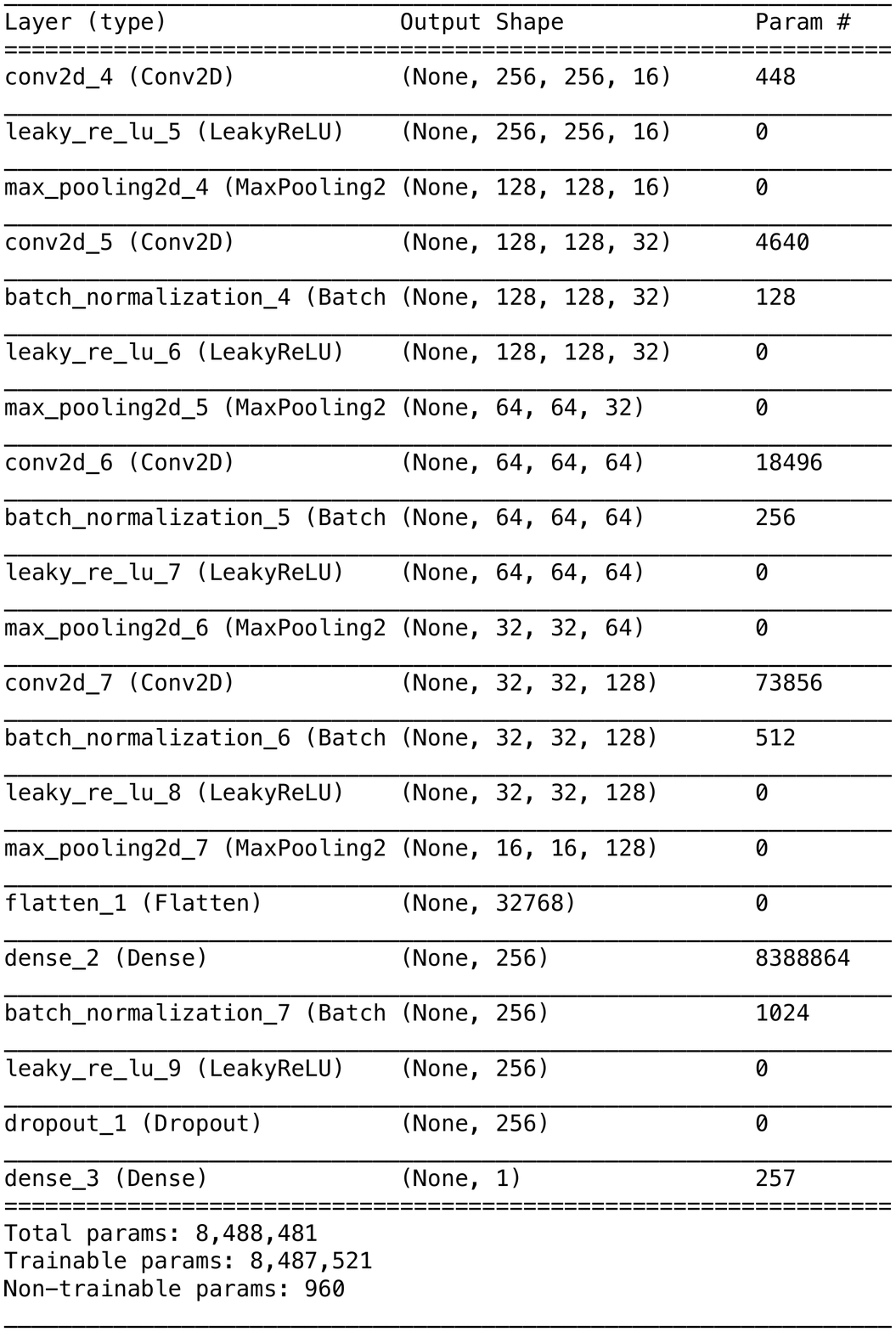}
\end{center}
   \caption{Description of the TeliNet architecture with Convolution layers and the tunable parameters.}
\label{telinetsummary}
\label{fig:onecol}
\end{figure}

It is a four-layer shallow Convolutional Neural Network. It consists of 2D convolutions, batch normalizations, LeakyReLU activation functions and max pooling layers. This architecture is represented in Figure~\ref{telinet} with its corresponding summary in Figure~\ref{telinetsummary}.

As seen, it is a very shallow network with less than 8.5 Million trainable parameters. That is much smaller than over 134 Million trainable parameters in VGGNet-16. The input images are resized from the original $512 \times 512$ pixels to $256 \times 256$ images and re-scaled to values between 0 and 1. 2D convolution is carried out on these images using 16 filters of size, $3 \times 3$ followed by LeakyReLU activations. After the initial convolutions, there are 3, 2D convolutions layers, each followed by batch normalization, LeakyReLU activation functions and maxpooling. The padding is same in all layers, with a unit stride for each convolution. The max pooling for each layers consists of $2 \times 2$ pixels with a stride of 2.

After the convolution layers, two dense layers, each of 256 units is used, followed by batch normalization and LeakyReLU activation functions. Unlike the convolutions layers, we use a 10\% dropout for this dense layer. Finally, we have a Sigmoid unit for binary classification as covid or non-covid.

Batch normalization smoothes the optimization problem by normalizing the inputs to each layer. This helps fight the covariant shift problem~\cite{sergey} and also enables a quicker and more stable training~\cite{NEURIPS2018_905056c1} of the data set using TeliNet. It leads to faster convergence and takes advantage of a wider range of learning rates. To further benefit from these features we use a learning rate scheduler that reduces the learning rate by a factor of 70\% with each epoch. This values was chosen after a series of pilot experiments. The advantages of using LeakyReLU as an activation function over other activation functions are well known~\cite{Xu}. It also facilitates a faster convergence, complimenting the speed-up due to fewer trainable parameters and batch normalizations.  

\subsection{VGGNet}
Simonyan and Zisserman~\cite{Simonyan15} presented the VGGNet architecture for a large scale image recognition. This work presented the use of the small $3 \times 3$ convolution filters that helped extend the depth of the deep convolution network to 16-19 network layers. Using this architecture Simonyan et al. were able to secure both the first and the second positions in the ImageNet~\cite{imagenet} challenge of 2014. 

\begin{figure}[t!]
\begin{center}
   \includegraphics[height=6in,width=1.2\linewidth]{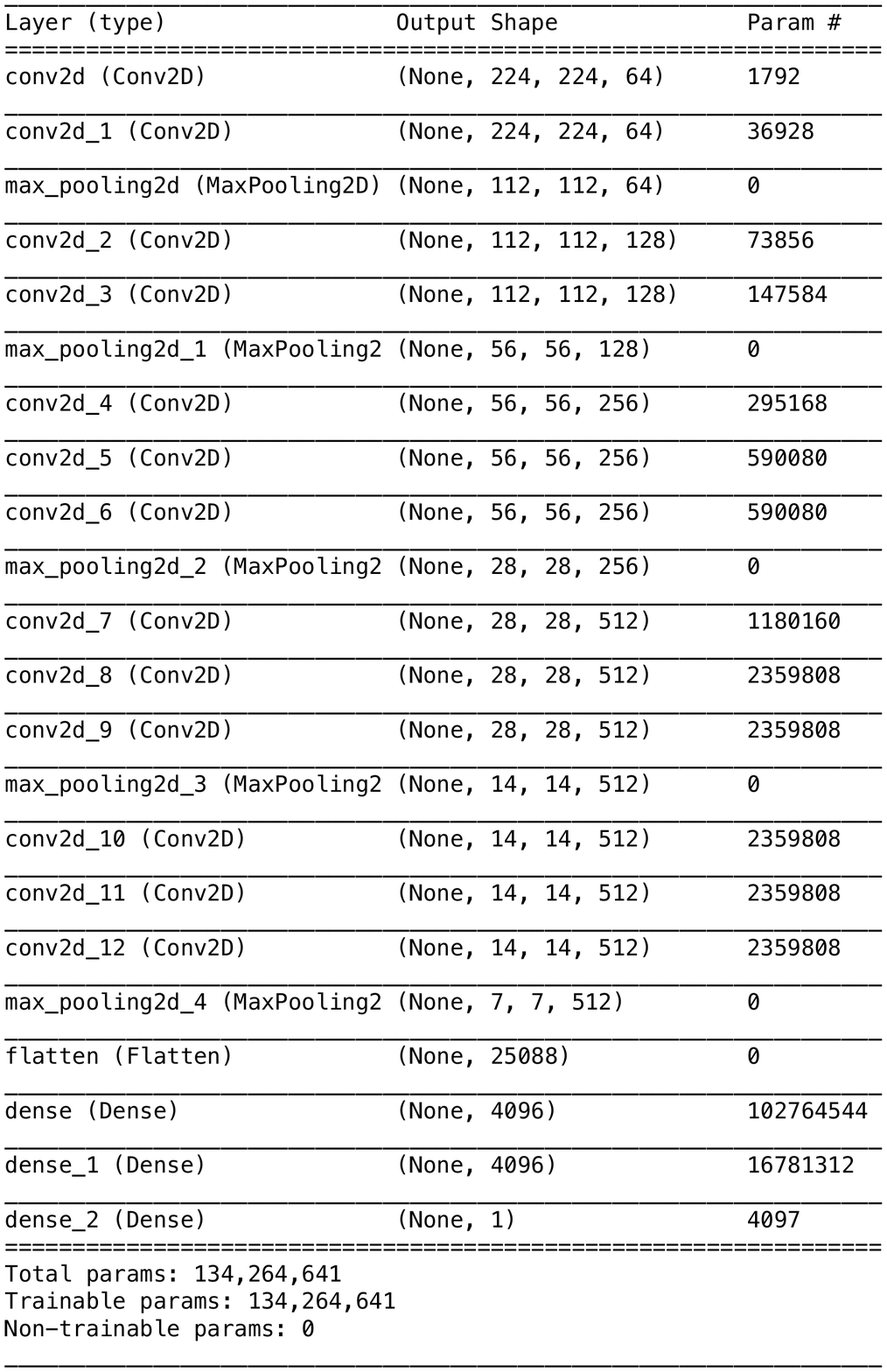}
\end{center}
   \caption{Description of the VGGNet-16 architecture with Convolution layers and the tunable parameters.}
\label{fig:vggnetarch}
\label{fig:onecol}
\end{figure}

For this research we used VGGNet-16 that consists of 16 layers and over 134 Million trainable parameters.
The architectural summary is presented in Figure~\ref{fig:vggnetarch}.
Each layer of convolutions involves filters of sizes $3\times 3$, ReLU activation function,and padding to retain the size with a stride of 1. The maxpooling in this architecture is a $2\times 2$ filter with a stride of 2. The simplicity of this approach is the uniformity of the filter sizes, padding and maxpooling across all layers.

The input to this architecture is a $224 \times 224$ pixel image convolved with 64 filters. Same convolutions are repeated in the next layer. This is followed by maxpooling and two layers of convolutions using 128 filters and maxpooling. Next, there are three convolutions using 256 filters and maxpooling. It is followed by 3 convolutions with 512 filters and maxpooling. The next two layers are 
fully connected layers with 4096 units each. The final layer is a Sigmoid unit for the binary classifcation into covid and non-covid images. For a detailed description of the VGGNet-16 architecture, please refer to the original work by Simonyan and Zisserman~\cite{Simonyan15}.

\section{Results}
We present a comparison of the results of TeliNet, VGGNet-16 and the benchmark approaches. In the training set there are 153681 covid slices and 181991 non-covid slices. The validation set contains 35016 covid slices and 40517 non-covid slices. 
Since the data sets are not balanced we use an F1 `macro' score metric to measure the efficacy of our approach just like the one used in the Benchmark paper~\cite{kollias2021mia}. These results are shown in Table 1.
\begin{table}
\begin{center}
\begin{tabular}{|l|c|}
\hline
Method &  F1 `macro' score \\
\hline\hline
benchmark~\cite{kollias2021mia} & 0.7 \\
VGGNet-16 & 0.72 \\
\textbf{TeliNet}(ours) & \textbf{0.81}\\
\hline
\end{tabular}
\end{center}
\label{restable}
\caption{Validation set F1 score results using the three approaches.}
\end{table}

The results of TeliNet and VGGNet-16 were obtained on the same hardware, comprising of a MacBook Pro, with a 2.3 GHz Dual-Core Intel Core i5 processor and 8 Gb of RAM. Both these approaches were trained using the RMSprop optimizer, an adaptive learning method, proposed by Geoffrey Hinton, after empircally comparing with the Adam optimizer~\cite{adam}. We also tried experiments using different sized batches ranging from 4 to 128 and finally settled on 32 for the final results due to better performance.

Since it is a binary classification problem, we used binary cross-entropy loss function to train these networks. Accuracy, precision and recall metrics were computed. Precision and recall were used to determine the F1 score by taking their harmonic mean.

Even though our solution has done better than the benchmark method and the VGGNet-16 architecture, there is a plenty of room for improvement. Our F1 `macro` score on validation data set is 81\% in comparison to 72\% and 71\% on the same data set by VGGNet-16 and the benchmark approach. On the test set TeliNet gives an F1 `macro' score of 70.86\% in comparison to 67\% for the benchmark score. The labels for the test set were not given to us and therefore these results are based on the outcomes of the experiments performed on the test sets by the competition organizing team.

It is important to note that these results are for 3D CT-scan series'. While it may seem that the reported results on the validation and the test sets have been reported on 2D slices, it is in fact a 3D series of these slices. Each of these series' provides a sequences of 2D slices. Our results were obtained on these series' with the label assigned based on the classification outcomes on these slices. Although our goal was to use a majority voting to determine the label for a series in these slices, it was seen that majority of these slices were classified to be belonging to a single class for each series anyway and so it was a very straightforward decision. 

\section{Conclusion}
In this research we introduce a shallow and lightweight but very effective CNN architecture, TeliNet, for binary classification of CT-scan series' of COVID-19 and non-COVID19 slices. We compare our results with VGGNet-16 and the benchmark results on this difficult data set. The results indicate that our approach does significantly better than the other two methods on the validation data set and better than the benchmark approach on the test set. The results on the test set using the VGGNet have not been evaluated since the test set labels were not provided to us.

Our contributions include, 1) a very simple CNN architecture trained to do better than more complex architectures, and 2) a lightweight architecture that can be run on a personal computer. These advantages make this architecture accessible to clinicians and researchers who may not always have a high end hardware.

However, this paper would not be complete without discussing the pitfalls of the Machine Learning research conducted in the diagnosis using chest X-rays and CT scan images. Roberts et al.~\cite{roberts21} conducted a review of 62 published papers after a thorough screening of 2,212 studies conducted between Jan. 1, 2020 to October 3, 2020 that presented new Machine Learning models for the diagnosis or prognosis of COVID-19 from chest X-ray and CT images. In each and every paper they found either methodological flaws and/or underlying biases and lack of reproducibility which render these proposed approaches useless of any potential clinical use. Keepng this in mind it is worthwhile to know that we have presented not only detailed  architectures and the methodologies applied, we also make our code available on Github\footnote{https://github.com/nayeemmz/TeliNet} for easy reproducibility. The dataset has been validated and the results of the competition have been uploaded to a Leaderboard for easy evaluation of the approaches and the validation of the results.



{\small
\bibliographystyle{ieee_fullname}
\bibliography{egbib}
}

\end{document}